\documentclass{article}
\usepackage[T1]{fontenc} 
\usepackage[utf8]{inputenc} 
\usepackage{ismir,amsmath,cite,url}
\usepackage{graphicx}
\usepackage{booktabs}
\usepackage{color}
\usepackage[table,xcdraw]{xcolor}
\usepackage{lineno}
\usepackage{multirow}
\usepackage{cite}
\usepackage{url}

\title{Music Discovery Dialogue Generation Using\\Human Intent Analysis and Large Language Models}

\multauthor
{SeungHeon Doh$^{\flat}$ \hspace{1cm} Keunwoo Choi$^{\natural}$ \hspace{1cm} Daeyong Kwon$^{\flat}$} { \bfseries{Taesu Kim$^{\sharp}$ \hspace{1cm} Juhan Nam$^{\flat}$ \hspace{1cm}}\\ 
$^{\flat}$ Graduate School of Culture Technology, KAIST, South Korea \\
$^{\natural}$ Prescient Design, Genentech, New York, NY, USA
\\
$^{\sharp}$ Department of Industrial Design, KAIST, South Korea\\
{\tt\small \{seungheondoh, ejmj63, tskind77, juhan.nam\}@kaist.ac.kr, choi.keunwoo@gene.com}
}
\def\authorname{S. Doh, K. Choi, D. Kwon, T.Kim and J. Nam}

\begin{document}

\maketitle
\begin{abstract}
A conversational music retrieval system can help users discover music that matches their preferences through dialogue. To achieve this, a conversational music retrieval system should seamlessly engage in multi-turn conversation by 1) understanding user queries and 2) responding with natural language and retrieved music. A straightforward solution would be a data-driven approach utilizing such conversation logs. However, few datasets are available for the research and are limited in terms of volume and quality. In this paper, we present a data generation framework for rich music discovery dialogue using a large language model (LLM) and user intents, system actions, and musical attributes. This is done by i) dialogue intent analysis using grounded theory, ii) generating attribute sequences via cascading database filtering, and iii) generating utterances using large language models. By applying this framework to the Million Song dataset, we create -- \textbf{LP-MusicDialog}, a \textbf{L}arge Language Model based \textbf{P}seudo \textbf{M}usic \textbf{D}ialogue dataset, containing over 288k music conversations using more than 319k music items. 
Our evaluation shows that the synthetic dataset is competitive with an existing, small human dialogue dataset in terms of dialogue consistency, item relevance, and naturalness. Furthermore, using the dataset, we train a conversational music retrieval model and show promising results.\footnote{Our code is available at https://github.com/seungheondoh/lp-music-dialog/}

\begin{figure}[!t]
\centering
\includegraphics[width=\linewidth]{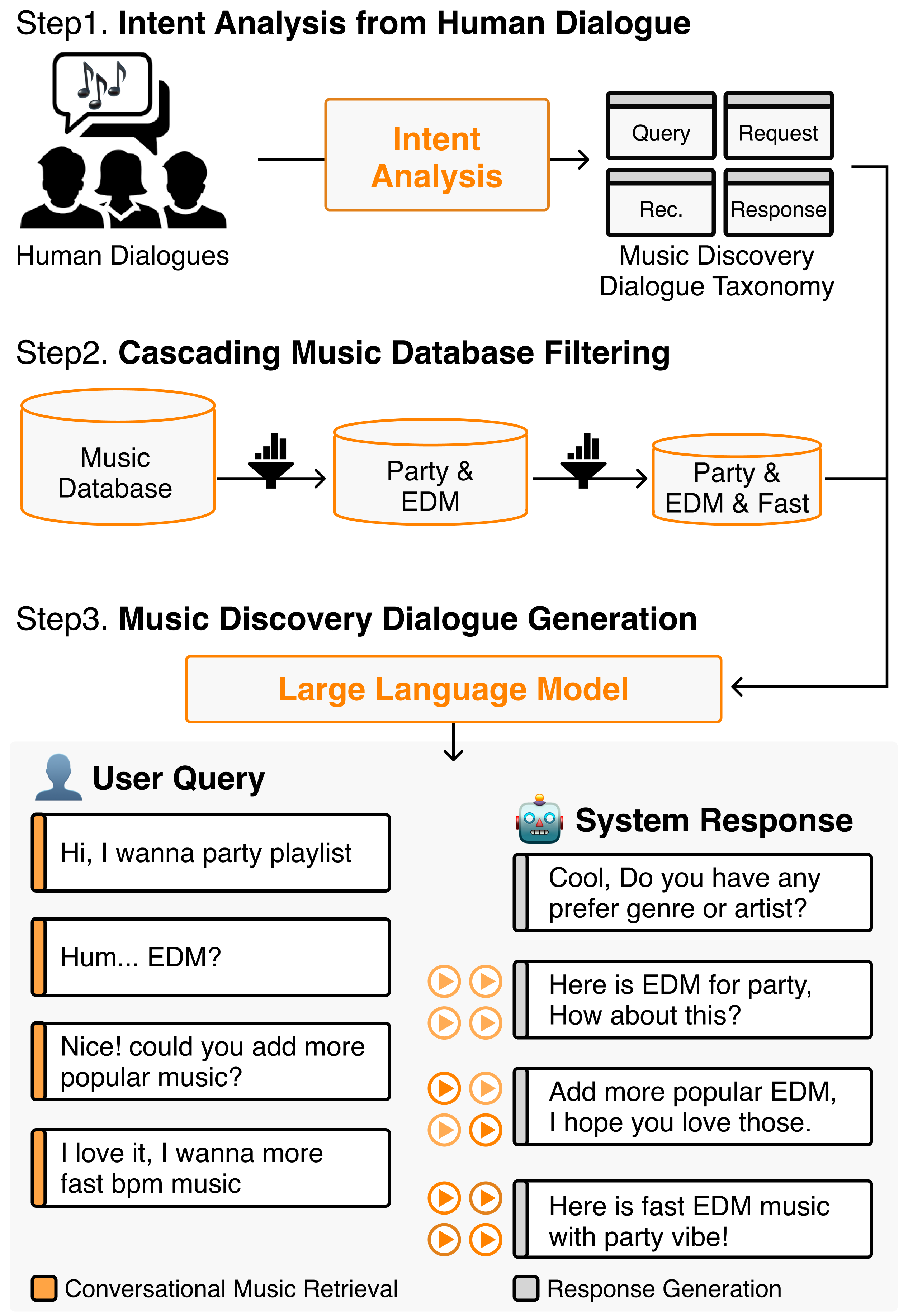}
\caption{The generation process of pseudo musical dialogue.}
\end{figure}

\end{abstract}

\section{Introduction}\label{sec:introduction}
In recent years, conversational systems have emerged as a promising solution to enhance user experience in various domains~\cite{christakopoulou2016towards, li2018towards,jannach2021survey, he2023large}, including conversational music retrieval and recommendation~\cite{chaganty2023beyond, leszczynski2023talk}. The goal of a conversational music system is to assist users in finding their desired music through dialogues.
Such a system should possess three key capabilities: i) to understand the \textit{intents} and \textit{musical needs} of users from their queries expressed in natural language, ii) to generate responses and facilitate human-like interaction, iii) to find music that aligns with the user's preferences by taking previous dialogues into account.

Currently, the primary challenge of developing a conversational music retrieval system is the scarcity of large-scale public datasets. Chaganty~et~al.~\cite{chaganty2023beyond} introduce the Conversational Playlist Curation Dataset (CPCD). This crowd-sourced dataset comprises human-to-human dialogues that simulate the process of music discovery. However, as it relies on a manual process, the dataset is small and exhibits biases from the music streaming platforms used by the recommenders. To address this problem, Leszczynsk et al.~\cite{leszczynski2023talk} propose a dialogue generation framework through random walks in the music-text joint embedding space and dialogue inpainting~\cite{dai2022dialog}. However, this approach requires a high-quality music-text joint embedding and needs to use template-based system responses as input. As a result, the system's responses are always composed of limited format utterances, leading to low naturalness in human evaluation.

In this paper, we introduce a framework for generating \textit{human-like} music discovery dialogues using intent and a large language model (LLM). The proposed framework is based on the existing method~\cite{leszczynski2023talk}, but we address their limitations by employing \textit{cascading music database filtering} instead of a joint embedding and extensive \textit{intent analysis} for naturalness. Using the grounded theory approach~\cite{glaser1968discovery}, we analyze a dataset of human music discovery dialogues and develop taxonomies for user intents, system actions, and musical attributes relevant to the task of music discovery. Furthermore, we introduce a model-free attribute sequence generation by applying cascading filtering to a multi-label music annotation database. Finally, we synthesize music discovery dialogues through an LLM using the created attribute sequences and human intents/actions.

Our contributions are threefold: First, we analyze music discovery dialogues and propose a taxonomy. Second, we introduce the LP-MusicDialog dataset, a large-scale synthetic dialogue dataset created using human intent and an LLM. Third, we present extensive objective and subjective evaluations to demonstrate the effectiveness of LLM-based pseudo-music dialogues.

\section{Related Work}

Recently, there has been some progress in conversational music systems including language-based music retrieval~\cite{manco2022contrastive,doh2023toward, doh2024enriching}. However, existing systems are often limited to single-turn tasks. This means users are not able to refine their queries to obtain a highly satisfactory outcome. Towards multi-turn dialogues, 
Chaganty et al.~\cite{chaganty2023beyond} released the Conversational Playlist Curation Dataset (CPCD), which comprises 917 dialogues averaging 5.7 turns each. Unlike single-turn retrieval, conversational retrieval takes into account previous chat history to find relevant items. The model in~\cite{chaganty2023beyond} aggregates the context of history embedding and the current query embedding using average pooling, then uses contrastive loss to maximize the similarity between them. However, their model showed limited performance due to the small size of CPCD.

A solution to the data scale issue is synthesizing data using existing datasets and language models. Recently, synthetic datasets that bridge natural languages and music have been proposed to enhance music understanding~\cite{liu2024music, gardner2023llark}, captioning~\cite{doh2023lp, gardner2023llark}, reasoning~\cite{gardner2023llark}, and retrieval~\cite{doh2024enriching}. For the conversational music retrieval, Leszczynski et al.~\cite{leszczynski2023talk} proposed a two-stage data synthesis framework: musical attribute sequence generation via random walk and utterance generation through dialogue inpainting~\cite{dai2022dialog}. The musical attribute sequence represents the evolution of user queries over turns (e.g., ask for workout music in the first turn, then refine the results to be also pop music). During utterance generation, a language model creates user queries using system responses as input, which include sampled musical attributes and a static template.\footnote{For example, ``Of course! Let me add some songs described as <musical attributes>. What else?"} As a result, they created one million multi-turn music discovery dialogues: TtW Music, leveraging a private playlist dataset. However, this approach encounters issues with model errors in the music-text joint embedding space used for the random walk method and faces challenges with response consistency due to the reliance on manually created templates for system responses.


Deep understanding in music query has to be preceded aforementioned data generation. 
So far, query understanding has primarily focused on describing musical needs. Downie and Cunningham~\cite{downie2002toward} analyzed 161 music queries and categorized them into 1)~information needs, 2)~desired outcomes, 3)~intended uses for the information, and 4)~social and contextual elements. Bainbridge et al.~\cite{bainbridge2003people} utilized the grounded theory approach to analyze 502 real-world music queries, expanding upon prior research with 10 types of need descriptions. Lee~\cite{lee2010analysis} analyzed 1,705 Google Answers queries to propose a refined taxonomy for information needs and searching behavior. Despite these efforts, previous studies have been limited to the information needs (musical attributes) contained in queries, and intent in multi-turn queries has not received significant attention.

\begin{table*}[!t]
\centering
\resizebox{\textwidth}{!}{%
\begin{tabular}{lllr}
\toprule
User Intent            & Description                                                 & Example                                 & \%    \\ \midrule
\multicolumn{4}{l}{{\color[HTML]{9B9B9B} \textit{Start Dialogue}}}   \\
Initial Query & User initiates the inquiry with a specific request. & ``Hi, I want to create a playlist for hiking." & 18.5 \\
Greeting                & User initiates the dialog, often with greeting words.   & ``Hello, I would like... / ``Good Morning! Let's start..."                & 12.7          \\ \midrule
\multicolumn{4}{l}{{\color[HTML]{9B9B9B} \textit{\text{Item Discovery Query (Retrieval / Recommendation)}}}}                                  \\ 
Positive Filter              & User requests to include an additional criterion.       & ``I would like to add a bit of Rihanna."     & \textbf{76.7} \\
Negative Filter           & User requests to negative criterion in this turn.    & ``I do prefer them to not have any lyrics" & 3.7           \\
Continue                & User requests for more songs with the current criteria. & ``Those are good songs. More like these would be great."                 & 6.8           \\ \midrule
\multicolumn{4}{l}{{\color[HTML]{9B9B9B} \textit{\text{Item Understanding Query (Question Answering)}}}}                                      \\
Item Attribute Question & User questions attributes of music.                     & ``Do you know where Samer (Artist) is from?"        & 0.2           \\ \midrule
\multicolumn{4}{l}{{\color[HTML]{9B9B9B} \textit{\text{Feedback Response}}}}                                                                  \\
Accept Response         & User responds positively to the recommendations.        & ``Thank you, they are perfect"             & \textbf{44.4} \\
Reject Response         & User responds negatively to the recommendations.        & ``I still didn't get any song suggestions..."           & 4.8           \\ 

\bottomrule \end{tabular}
}
\vspace{-3mm}
\caption{Taxonomy for user intents. \textbf{\%} represents the percentage of occurrences within the user query.}
\label{tab:user}
\end{table*}

\begin{table*}[!t]
\centering
\resizebox{\textwidth}{!}{%
\begin{tabular}{lllr}
\toprule
System Action      & Description                                                            & Example                                                         & \% \\ \midrule
\multicolumn{4}{l}{{\color[HTML]{9B9B9B} \textit{Request}}}                                                                                                                \\
Feedback Request   & System requests the user to evaluate recommendations.              & ``What about these ones?" & 12.8       \\
Detail Attribute Request &
  System requests for the user's needs or desires for recommendations. &
  ``Are there any particular artists you want to see?" &
  20.8 \\ \midrule
\multicolumn{4}{l}{{\color[HTML]{9B9B9B} \textit{Item Discovery Response (Retrieval / Recommendation)}}}                                                                   \\
Passive Recommendation  & System recommends music based on user's preferences.               & ``Here are some pop songs for kids." & \textbf{71.4}       \\
Active Recommendation   & System proactively recommends music without being explicitly asked. & ``Nice! I added a couple more."           & 14.9      \\ \midrule
\multicolumn{4}{l}{{\color[HTML]{9B9B9B} \textit{Item Understanding Response (Question Answering)}}}                                                                       \\
Item Attribute Answer &
  System answers to user's question with musical attributes. &
  ``Frederic Chopin was a Polish composer..." &
  0.1 \\ \midrule
\multicolumn{4}{l}{{\color[HTML]{9B9B9B} \textit{General Response}}}                                                                                                       \\
Parroting Response & System responds to the user's inquiry by mirorring.                & ``Here's some picks from The Who." & 25.4         \\
Sympathetic Response &
  System responds to the user's inquiry with human-like sympathy. &
  ``Excellent! Looks like a great Kickstart!"  &
  \textbf{57.2} 
\\ \bottomrule
\end{tabular}
}
\vspace{-4mm}
\caption{Taxonomy for system actions. \textbf{\%} represents the percentage of occurrences within the system response.}
\vspace{-3mm}
\label{tab:sys}
\end{table*}

\section{Dialogue Intent Analysis}\label{sec:intent_analysis}

\subsection{Taxonomy Development}
For the dialogue-specific music discovery taxonomy, we analyze the existing human-to-human music dialogue dataset (CPCD~\cite{chaganty2023beyond}) using the grounded theory approach~\cite{glaser1968discovery}. Grounded theory is a qualitative approach that creates refined theory from unstructured real-world data. In detail, (1) we adopt the taxonomy from previous research as our initial taxonomy. For user intent and system action, we use the \textit{conversational movie intent} taxonomy~\cite{cai2020predicting}, and for musical needs, we use \textit{music feature} taxonomy~\cite{lee2010analysis} as the initial taxonomy. (2) Three authors annotate ten dialogues using the initial taxonomy and discuss the limitations of the existing taxonomy. (3) We update the taxonomy and annotate a new randomly sampled five dialogues. This cycle of proposing, refining, and annotating was completed three times to ensure our taxonomy could capture the full range of scenarios present in the dialogue samples.


\subsection{Analysis Results}\label{sec:intent_analysis_results}

\subsubsection{Taxonomy for User Intents}
In Table~\ref{tab:user}, we categorize user intents into four main categories, with eight detailed sub-intents. The \textbf{Initial Query} is part of the `start-dialogue' category, serving as the kickoff point for recommendation dialogues. At times, users start interactions with \textbf{Greetings} to initiate the system into a more engaging mode of communication. The `item-discovery-query' captures user preferences for music retrieval and recommendations, subdivided into \textbf{Positive Filter} for add preferences, \textbf{Negative Filter} for discarding existing ones, or \textbf{Continue} to sustain the current preferences. \textbf{Item Attribute Question}, where users inquire about the precise attributes of music tracks, such as their genre, mood, tempo, and key/mode. Feedback responses are outlined, with users either expressing satisfaction (\textbf{Accept Response}) or dissatisfaction (\textbf{Reject Response}) to the recommended music.

\begin{table*}[!t]
\centering
\resizebox{\textwidth}{!}{%
\begin{tabular}{lllr}
\toprule
Musical Attribute      & Description                                           & Example                                         & \%         \\ \midrule
\multicolumn{4}{l}{{\color[HTML]{9B9B9B} \textit{Metadata}}}                                                                             \\
Track          & A single musical work, recording, performance              & Can you add \textbf{montero from lil nas}?                & 4.5           \\
Artist         & A creator or performer of music.                      & How about some more \textbf{Justin Timberlake}?          & \textbf{43.1} \\
Year           & The time of music's creation or release               & Could you throw in some \textbf{90s} hip hop?            & 6.6           \\
Popularity     & The widespread acclaim of music.                      & Awesome! How about more male disco \textbf{hits}?        & 0.7           \\
Culture        & The national or regional influences on music.         & i like \textbf{Nigerian} songs                           & 0.9           \\ \midrule
\multicolumn{4}{l}{{\color[HTML]{9B9B9B} \textit{Similar with Music Entity}}}                                                            \\
Similar Track  & The song is similar to the specific track.                 & Can I have more songs \textbf{similar to Jessie's Girl}? & 1.1           \\
Similar Artist & The song is similar to the specific artist.                & \textbf{Adele and Sia type} of music                     & 5.0           \\ \midrule
\multicolumn{4}{l}{{\color[HTML]{9B9B9B} \textit{User \& Listening Context}}}                                                            \\
User           & The listeners characterized by demographics.          & I want to create a fun playlist for the \textbf{kids}    & 1.9           \\
Theme          & A context for music listening related to location, time, usage, and activity.                     & hi please i need a playlist to \textbf{workout} with     & 15.2          \\
Mood           & The emotional tone conveyed by music, or the emotional state of user.       & I want to create playlist for when I'm \textbf{sad}      & 6.7           \\ \midrule
\multicolumn{4}{l}{{\color[HTML]{9B9B9B} \textit{Music Content Information}}}                                                            \\
Genre & A category of music characterized in form, style, or subject matter.      & I would love to start with \textbf{classical} music & \textbf{17.2} \\
Instrument     & A tool or device designed to produce musical sounds   & What other \textbf{quartet music} do you suggest?        & 1.1           \\
Vocal & The singing voice in music, including styles, techniques, and expressions & try some \textbf{female vocalists} too              & 1.2           \\
Tempo          & The speed or pace at which a piece of music is played & Add me some \textbf{slow slow songs} from juice wrld     & 1.3           \\
Key / Mode          & The tonal information of music                        & (Not appear)           & 0.0       \\ \bottomrule   
\end{tabular}
}
\vspace{-3mm}
\caption{Taxonomy for musical attributes. \textbf{\%} represents the percentage of occurrences within the user query.}
\label{tab:music}
\vspace{-3mm}
\end{table*}

\subsubsection{Taxonomy for System Actions}
Table~\ref{tab:sys} shows the system action taxonomy, structured into four primary categories, and seven specific intents. The `request' category enhances the search experience, either through post-recommendation \textbf{Feedback Request} to gauge satisfaction or \textbf{Detail Attribute Request} to clarify vague or incomplete user queries. To address user queries for music discovery, the system can adopt a \textbf{Passive Recommendation} approach to comply with user requests. This approach is similar to a retrieval task because there is an explicit query from the user. The system also proactively engage through \textbf{Active Recommendation}. It is similar to a recommendation task because it implicitly uses the context of the dialogue even without an explicit query from the user. 

In \textbf{Item Attribute Answer}, the system responds to users' inquiries about specific musical attribute questions such as genre, mood, tempo, and key/mode. For general responses, the system mimics user requests in its recommendations (\textbf{Parroting Response}) to affirm that user preferences are being considered, or it may adopt a more empathetic stance (\textbf{Sympathetic Response}) to foster a more human-like interaction.

\subsubsection{Taxonomy for Musical Attribute}
\vspace{-1mm}
Musical attributes such as genre, mood, and artist are closely related to user preferences. They are categorized into (1) objective metadata produced when a track is registered on the platform, (2) subjective similarity with music entities, (3) user \& listening context, and (4) music content information (Table~\ref{tab:music}). Metadata is mainly associated with entity recognition~\cite{epure2023human}, where \textbf{Track} refers to requests for a single music recording entity, \textbf{Artist} denotes user requests for tracks released by a specific artist, \textbf{Year} reflects the era/year in which a piece of music was released, and \textbf{Popularity} indicates the level of attention a piece of music has received. \textbf{Culture} is related to the national or regional style of the music, often linked to the artist's nationality. Unlike objective metadata, \textbf{Similar Track} and \textbf{Similar Artist} are subjective musical attributes, representing connections to other track or artist entities. The user \% listening context consist of \textbf{User}, which is related to the demographics of the listener, and \text{Theme}, which encompasses the location, time, usage, and activities associated with listening. \textbf{Mood} refers to the emotional tone conveyed by the music or the listener's emotional state. Lastly, the categories tied closely to music content itself include \text{Genre}, which relates to high-level music form and style, and aspects associated with timbre such as \textbf{Instrument} and \textbf{Vocal}, as well as \textbf{Tempo} related to rhythm, and \textbf{Key/Mode} for tonal information. 


\subsubsection{Dialogue intent annotation}
After the development of the taxonomy, the three annotating authors proceeded to annotate user intent, system action, and musical attributes according to the proposed taxonomy within a new sampled 30 CPCD~\cite{chaganty2023beyond} dialogues. To measure the level of agreement among the annotators for the degree of concurrence, we employed Krippendorff's alpha~\cite{krippendorff2009testing}. The results showed high agreement levels, with Krippendorff's alpha scores of 0.83 for user intent, 0.85 for system action, and 0.71 for musical attributes. Following this agreement phase, each annotator independently annotated a portion of the total 888 CPCD dialogues, excluding 29 error samples out of the total 917 dialogues.

\subsection{Intent Analysis and Findings}
The proportion of categories in each taxonomy is on the rightmost columns of Tables 1 -- 3. For user intent (in Table 1), the majority of item discovery queries progress through the \textit{Positive Filter} (76.7\%), whereas the opposite, \textit{Negative Filter}, constitute a smaller portion (3.7\%). \textit{Item Attribute Questions} are almost non-existent (0.2\%). The majority of recommendation are accepted (44.4\% of \textit{Accept Response} vs 4.8\% of \textit{Reject Response}). 
Regarding system actions, a notable observation is that the system predominantly provides recommendations passively, rather than actively offering suggestions (71.4\% vs 14.9\%). 
In the case of general responses, the system favors eliciting human-like conversations through sympathetic responses over merely parroting back information (57.2\% vs 25.4\%). In the case of music, categories such as artist (43.1\%), genre (17.2\%), and theme (15.2\%) occupy a significant portion of the user queries. This may suggest that in a \textit{Wizard of Oz} setting~\cite{dinan2018wizard}, where the recommender utilizes music streaming platforms, the platform's search support may be limited to these categories. Conversely, similarity queries or music content queries beyond genre are rarely employed.

\section{Music Discovery Dialogue Generation}\label{sec:related_works}
In this section, we describe the framework for generating music discovery dialogue consisting (1) attribute sequence generation and (2) utterance generation. This two-step approach is an improved version of previous research~\cite{leszczynski2023talk} with model-free attribute sequence generation by \textit{Cascading music database filtering}, and utterance generation by a LLM and \textit{Intent Analysis}.

\subsection{Musical Attribute Sequence Generation}\label{sec:cascading}
\begin{figure}[!t]
\centering
\includegraphics[width=\linewidth]{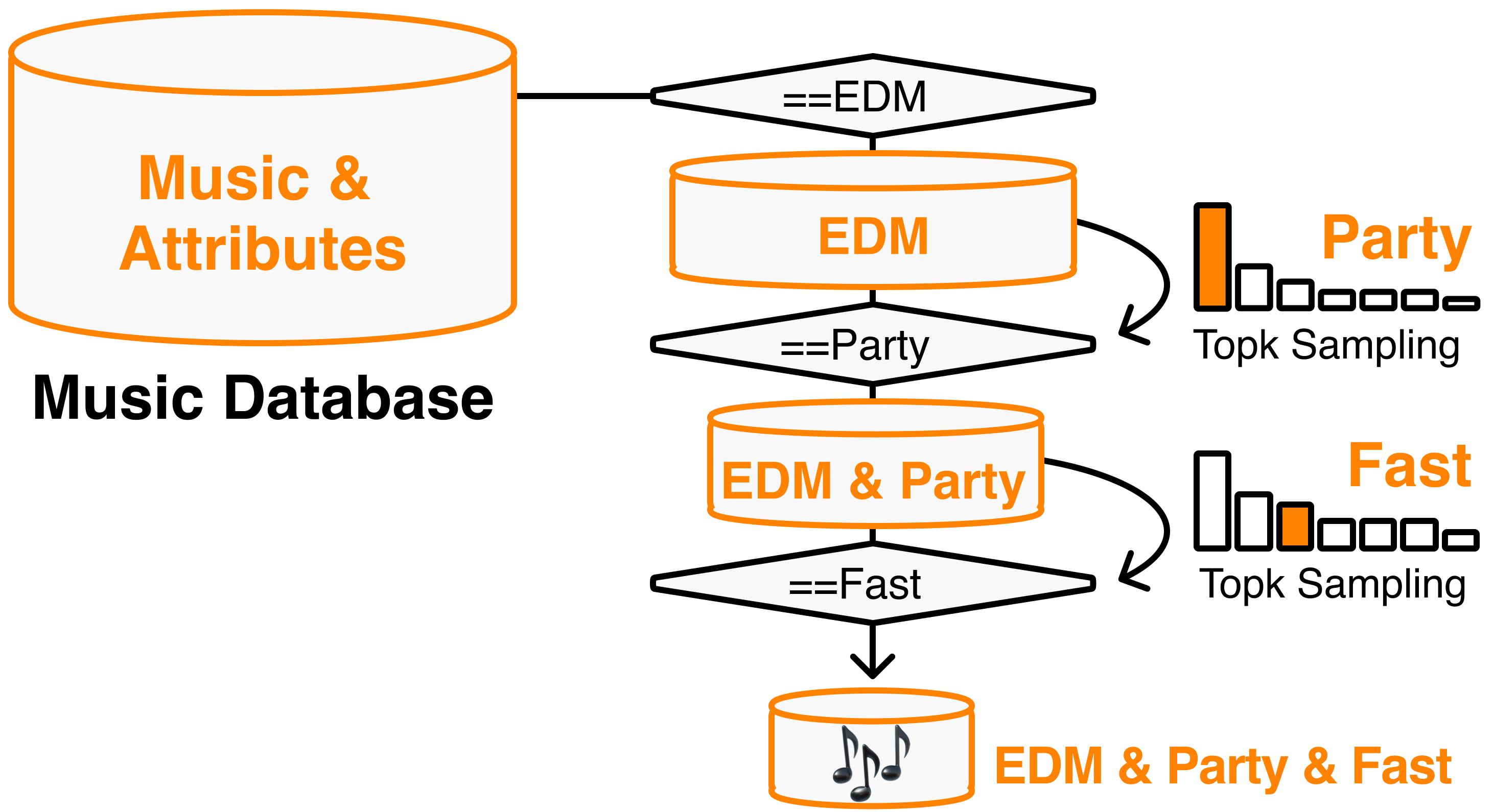}
\vspace{-3mm}
\caption{Cascading database filtering.}
\label{fig:cas}
\vspace{-3mm}
\end{figure}

For attribute sequence generation, we employ a series of cascading filters on a multi-label annotated dataset to extract samples with overlapping semantics. Inspired by \cite{johnson2017clevr}, which utilized \textit{functional programs} to construct a visual reasoning dataset, we apply this approach to the domain of music discovery dialogue using user intents such as add filter, remove filter, and continue. Figure~\ref{fig:cas} illustrates an example of a cascading data filter. For example, a user may initially request songs in the \textit{EDM} genre, narrow it down to a \textit{party} theme, and finally specify a need for \textit{fast tempo} music. This sequence can be represented by a series of connected functional program filters: \textit{filter}(tempo:fast, \textit{filter}(theme:party, \textit{filter}(genre:edm, database))). This method requires the types of filters and musical attributes. We derive filter types from the annotated intents~(Section~\ref{sec:intent_analysis_results}). For musical attributes, we initially conducted random sampling. Subsequently, to ensure diverse sampling with high co-occurrence with previous attributes, we employed top-$k$ sampling ($k$=20), which involves randomly selecting a word from the top $k$ words with the highest frequency. During top-$k$ sampling, we include attributes from the \textit{metadata} and \textit{similar with music entity} categories.

\subsection{Utterance Generation via Language Model}
The sequence of musical attributes and annotated intent becomes prompts for LLMs to generate user and system utterances. Our proposed utterance generation follows the formulation: $x_{text} = f_{\text{LLM}}(\mathcal{P}, y_{intent},y_{music})$, where $y_{intent}$ and $y_{music}$ refer to the annotated intent/action (Sec \ref{sec:intent_analysis_results}) and the sampled musical attribute (Sec \ref{sec:cascading}), respectively, and $\mathcal{P}$ is the prompt for dialogue generation. We assumed that dialogue intent and musical attribute can serve as effective conditions for human-like dialogue generation. 

\section{Dataset: LP-M\lowercase{usic}D\lowercase{ialog}}

\subsection{Data Source}
To construct our synthetic dialogue dataset, we utilize the Million Song Dataset (MSD)~\cite{bertin2011million}, which has rich metadata including track details, artist information, release year, and artist familiarity. We quantize year and artist familiarity into decades and popularity, respectively.\footnote{We categorize the top 10\% of artist familiarity as high popularity, 10-30\% as mid popularity, and the lower 30\% as low popularity.} To cover a wide range of musical attributes as listed in Table~\ref{tab:music}, we interlink multiple annotation datasets using track IDs. For culture, mood, theme, genre, instrument, and vocals, we utilize tags from Tagtraum~\cite{schreiber2015improving}, Last.fm~\cite{won2020multimodal, won2021transformer}, and AllMusic~\cite{schindler2019multi}.\footnote{As the `style' category in AllMusic is structured as sub-genres, we merged it with `genre' category.} For similar track attribute, we incorporate merged data from the Art of the Mix playlist~\cite{mcfee2012hypergraph} and EchoNest taste profiles. 
We utilize a weighted matrix factorization technique~\cite{hu2008collaborative} to create a similarity matrix of item vectors. We then annotate the top-$k$ similar tracks for each track ($k$=128). For artist similarity, we use cultural similarity annotation from the OLGA~\cite{korzeniowski2021artist} dataset. Finally, for key/mode and tempo, we extract beats per minute (BPM), 24 key/mode using the pretrained classifier~\cite{bock2019multi, korzeniowski2018genre}.\footnote{BPM was quantized into three text label: songs below 70 BPM were classified as slow, those between 70 BPM and 130 BPM as moderate, and those above 130 BPM as fast.}

\subsection{Creation Process}
Based on the proposed pseudo dialogue generation method, we created LP-MusicDialog, an LLM-based Pseudo Music Dialogue dataset. We integrate user intents and system actions annotated in CPCD dialogues (Sec.\ref{sec:intent_analysis_results}) with musical attributes obtained through cascading music database filtering (Sec.\ref{sec:cascading}) as inputs for GPT~3.5-turbo. 
At each cascading filtering step, we randomly sample 10 tracks to link with the dialogue turn. As a result, we acquire 287,675 {user query, system response, music item} triplets for each turn of the dialogue.

\begin{figure*}[!t]
\centering
\includegraphics[width=\linewidth]{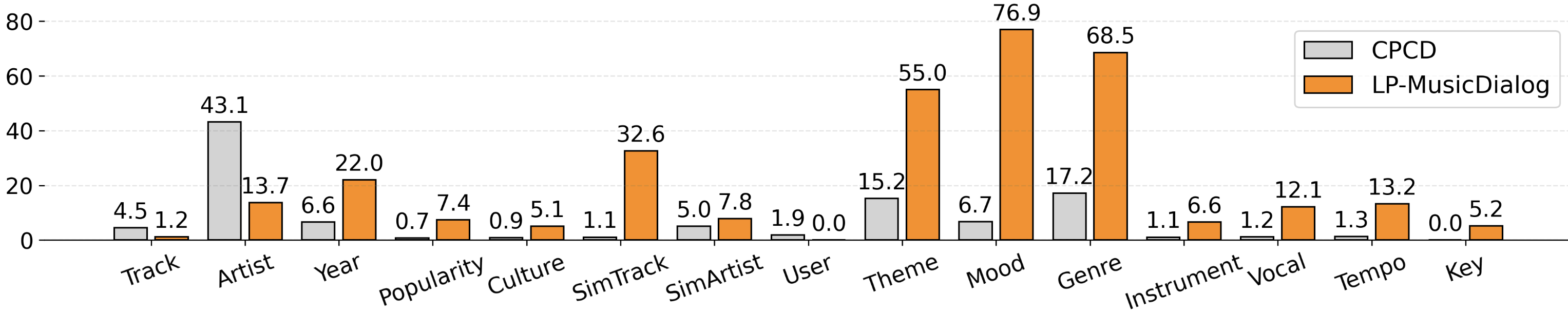}
\vspace{-7mm}
\caption{The ratios of musical attributes in music discovery dialogues.}
\vspace{-3mm}
\label{fig:attr}
\end{figure*}

As detailed in Table~\ref{tag:statistics}, LP-MusicDialog is significantly larger and more diverse than existing datasets. Compared to the human dialogue dataset CPCD~\cite{chaganty2023beyond}, it contains $\times$313 larger dialogues, a more than $\times$10 diverse vocabulary, and nearly four times more tracks. While remaining open to the public, our dataset is on par with a private dataset, TtW~\cite{leszczynski2023talk}, in many aspects, with plans to expand further upon confirming active usage. In contrast, LP-MusicDialog is not only publicly available but also offers an extensive collection of connected tracks. Figure~\ref{fig:attr} shows the musical attribute ratio in dialogue. Unlike CPCD~\cite{chaganty2023beyond}, where a significant portion is occupied by artist and genre due to platform bias, LP-MusicDialog shows a higher proportion of dialogues concerning music content and user listening context. Although the proportion of queries for entity recognition such as track and artist has decreased, the percentage of similarity queries has increased. 

\begin{table}[!t]
\centering
\resizebox{\linewidth}{!}{%
\begin{tabular}{lrl}
\toprule
Data Source & \# of Track & Musical Attributes                      \\ \midrule
Million Song Dataset~\cite{bertin2011million}         & 1,000,000 & Track,Artist,Year,Popularity \\ \midrule
TagTraum~\cite{schreiber2015improving} & 280,831 & \multirow{3}{*}{\begin{tabular}[c]{@{}l@{}}Genre, Instrument, Vocal, \\ Mood, Theme, Culture \end{tabular}} \\
Last.fm~\cite{won2020multimodal,won2021transformer}      & 344,865 &                              \\ 
AllMusic~\cite{schindler2019multi}     & 507,435 &                              \\ \midrule
Art of the Mix~\cite{mcfee2012hypergraph}         & 119,686 & \multirow{2}{*}{Similar Track}                \\
TasteProfile~\cite{bertin2011million} & 380,462 &                 \\ \midrule
OLGA~\cite{korzeniowski2021artist}         & 542,364 & Similar Artist               \\ \midrule
Madmom Key/Mode~\cite{korzeniowski2018genre}   & 992,525 & Key/Mode                          \\
Madmom Tempo~\cite{bock2019multi} & 978,759 & Tempo      \\ \bottomrule                 
\end{tabular}
}
\vspace{-3mm}
\caption{Data sources for the dialogue generation}
\vspace{-1mm}
\end{table}

\begin{table}[!t]
\centering
\resizebox{\linewidth}{!}{%
\begin{tabular}{lrrr}
\toprule
                          & CPCD~\cite{chaganty2023beyond}    & TtW~\cite{leszczynski2023talk} & LPMD~\textbf{(Ours)} \\ \midrule
\# of Dialog              & 917     & 1,037,701 & 287,675        \\
\# of Tracks              & 106,736 & 332,594   & 391,465        \\
\# of Vocab               & 9872    & \color{gray}{N/A}       & 105,832        \\
Avg.\# of turns          & 5.7     & 5.6       & 4.97           \\
Avg. query len.            & 54.4    & 80.3      & 63.8           \\
Avg. response len.        & 45.8    & \color{gray}{N/A}       & 87.8          \\
Public Available          & Yes       &   No       & Yes            \\ \bottomrule
\end{tabular}
}
\vspace{-2mm}
\caption{Statistics of conversational music retrieval datasets. LPMD stands for \textbf{LP}-\textbf{M}usic\textbf{D}ialog.}
\label{tag:statistics}
\vspace{-2mm}
\end{table}

\subsection{Human Evaluation}
Following previous work~\cite{leszczynski2023talk}, we assess the quality of our generated data through human evaluation, focusing on three key aspects: 1) Consistency - evaluating if the user preferences are coherent across dialogue turns; 2) Relevance - determining the alignment between the retrieved music items and the user query; 3) Naturalness - assessing the likelihood of such a conversation occurring in real life. Unlike previous work, we adhere to mean opinion score (MOS) that uses a 5-point Likert scale instead of a 3-point scale. A total of 26 raters evaluated 10 randomly sampled dialogues each. Within 260 total ratings, we only reported dialogues assessed by three or more raters. 

\begin{table}[!t]
\centering
\resizebox{\linewidth}{!}{%
\begin{tabular}{lccc}
\toprule
                  & Consistency & Relevance  & Naturalness   \\ \midrule
LP-MusicDialog    &  3.79 ± \small{0.56}      &   4.04 ± \small{0.74}     &  3.87 ± \small{0.57}  \\
\hspace{2mm}+ Intent / Action &  3.88 ± \small{0.87} & 4.05 ± \small{0.53} & 4.24 ± \small{0.38}  \\ \midrule
Human Dialogue~\cite{chaganty2023beyond} & 3.90 ± \small{0.70} & 4.16 ± \small{0.68} & 4.29 ± \small{0.43}
\\ \bottomrule
\end{tabular}
}
\vspace{-3mm}
\caption{Mean opinion scores of the generated dialogues (LP-MusicDialog) and human dialogues~(CPCD~\cite{chaganty2023beyond}) }
\vspace{-3mm}
\label{tab:mos}
\end{table}

Table~\ref{tab:mos} presents the MOS evaluations for dialogues from CPCD~\cite{chaganty2023beyond} and LP-MusicDialog. To understand the impact of user intent and system action, we conducted an ablation study synthesizing dialogues with only musical attributes and prompts. Comparing the first and second rows, we find minimal differences in consistency and relevance, as both sets of generated dialogues utilize identical musical attributes. However, a notable distinction arises in naturalness, suggesting that LLMs can foster more human-like dialogue synthesis by incorporating intents and actions. In comparing human dialogues with our generated dialogues, we found that the generated dialogues perform comparably to the human dataset across all three metrics, within the standard deviation.

\section{Conversational Music Retrieval}\label{sec:conversation}
In this section, we present a benchmark of conversational music retrieval models. Unlike prior studies~\cite{chaganty2023beyond,leszczynski2023talk} that have music embeddings solely relying on metadata-based text modality,\footnote{\{title\} by \{artist\} from \{album\}} we expand it to include audio modality. We use a pre-train audio-text joint embedding model (TTMR++~\cite{doh2024enriching}), that consists of a text encoder that handles user queries and system responses and an audio encoder that takes music tracks. 
We freeze TTMR++ and add a trainable MLP layer for both text and audio encoder. To handle chat history, we use a chat embedding created by average-pooling the current query, previous queries, responses, and music embeddings. Two encoders are trained to maximize the cosine similarity between the chat embedding and target music embedding using the InfoNCE~\cite{oord2018representation} loss. In the inference stage, we extract chat embeddings for each turn in the same way as in the training stage and measure the similarity score with all music embeddings in the train and test splits. Based on the similarity score, we retrieve the most similar $k$ items by nearest neighbor search.

\begin{table}[!t]
\centering
\resizebox{\linewidth}{!}{%
\begin{tabular}{llccc}
\toprule
Model         & Type & Hit@10 & Hit@20 & Hit@100 \\ \midrule
BM25~\cite{robertson1994m}          & -     & 0.180  & 0.251  & 0.433   \\
Contriever~\cite{izacard2021unsupervised}    & Zeroshot    & 0.176  & 0.255  & 0.344   \\ \midrule
TTMR++~\cite{doh2024enriching}       & Zeroshot    & 0.201  & 0.275  & 0.505   \\
\hspace{2mm}+ LPMD only   & Finetune$_{\text{OD}}$    &   0.173     &   0.253     &  0.479       \\
\hspace{2mm}+ CPCD only   &  Finetune$_{\text{ID}}$    & 0.209  &  0.295 & 0.530   \\
\hspace{2mm}+ LPMD \& CPCD & Finetune$_{\text{OD+ID}}$    &   \textbf{0.219}    &   \textbf{0.304}     &  \textbf{0.533}      \\ \bottomrule
\end{tabular}
}
\vspace{-3mm}
\caption{Conversational music retrieval performance on CPCD Dataset. \textbf{OD} stands for out of domain. \textbf{ID} stands for in domain.}
\vspace{-3mm}
\label{tab:per}
\end{table}

We chose the CPCD dataset as the compared dataset and report Hit@$K$ as evaluation metric ($k$=\{10,20,100\}).\footnote{Excluding the lost audio due to YouTube crawling, we use 98,738 out of 106,736 tracks for evaluation. We use the official evaluation codebase: https://github.com/google-research-datasets/cpcd} Our baseline models are as follows: (1) BM25~\cite{robertson1994m}, as a sparse retrieval baseline; (2) Contriever~\cite{izacard2021unsupervised}, an unsupervised dense retrieval baseline; and (3) TTMR++~\cite{doh2024enriching}, an audio-text joint embedding baseline. Table~\ref{tab:per} shows the performance of conversational retrieval. Among the baselines, TTMR++ shows superior performance over BM25 and Contriever, highlighting the importance of the audio modality in the music domain. Furthermore, training TTMR++ with only LP-MusicDialog (i.e., inter-dataset evaluation) somewhat leads to a performance decrease. This is presumably due to the musical entity difference between the two datasets. Specifically, the LP-MusicDialog dataset derives from the MSD, which contains music up to the year 2010, while the CPCD dataset includes music extending up to the year 2023. Nonetheless, training with both LP-MusicDialog and CPCD results in performance improvement over training only with CPCD, suggesting the usefulness of the proposed dataset.

\section{Conclusion}
We proposed a musical dialogue generation approach with i) dialogue intent analysis using the grounded theory, ii) generating attribute sequences via cascading database filtering, and iii) generating utterances using a large language model. Our intent analysis underpins the synthesis of human-like conversations, demonstrating strengths in naturalness. Cascading filtering allows us to utilize music attributes from external music databases to generate new dialogues. The outcome, the proposed \textbf{LP-MusicDialog} covers a broader range of musical attributes and aids in the conversational music retrieval task. 

However, the proposed methods have several limitations. The first is that cascading filtering is sensitive to annotation errors~\cite{choi2018effects}. The second is that top-k sampling, by following the tag distribution, inevitably leads to a data imbalance problem. We hope that these limitations of cascading filtering will be addressed in future research by incorporating balanced sampling.








\bibliography{ISMIRtemplate}

%
%
%
%
%

\end{document}